\documentclass[12pt]{amsart}

\newcommand{\Rset}{\mathbb{R}}
\newcommand{\bP}{\mathbb{P}}
\newcommand{\SL}{\operatorname{SL}}

\newcommand{\X}{\mathrm{X}}

\newcommand{\cD}{\mathcal{D}}
\newcommand{\cG}{\mathcal{G}}
\newcommand{\cE}{\mathcal{E}}

\newcommand{\cP}{\mathcal{P}}
\newcommand{\lp}{\left(}
\newcommand{\rp}{\right)}

\newcommand{\Sym}{\operatorname{Sym}}

\newcommand{\hp}{{\hat{p}}}
\newcommand{\hq}{{\hat{q}}}
\newcommand{\hr}{{\hat{r}}}

\newcommand{\hy}{{\hat{y}}}
\newcommand{\hP}{{\hat{P}}}
\newcommand{\hT}{{\hat{T}}}

\newcommand{\hx}{{\hat{x}}}
\newcommand{\hPab}{{\hat{P}^{(\alpha,\beta)}}}
\newcommand{\hLa}{{\hat{L}^{(\alpha)}}}

\newcommand{\cf}{\mathtt{C}}
\newtheorem{thm}{Theorem}[section]
\newtheorem{prop}[thm]{Proposition}
\newtheorem{lem}[thm]{Lemma}

\newtheorem{definition}[thm]{Definition}

\begin{document}
\title[An extension of Bochner's problem]{An extension of Bochner's problem: exceptional invariant
   subspaces}
\author{David G\'omez-Ullate}
\address{ Departamento de F\'isica Te\'orica II, Universidad Complutense de Madrid, 28040 Madrid, Spain}
\author{ Niky Kamran }
\address{Department of Mathematics and Statistics, McGill University
Montreal, QC, H3A 2K6, Canada}
\author{Robert Milson}
\address{Department of Mathematics and Statistics, Dalhousie University, Halifax, NS, B3H 3J5, Canada}
\begin{abstract}
  A classical result due to Bochner characterizes the classical
  orthogonal polynomial systems as solutions of a second-order
  eigenvalue equation. We extend Bochner's result by dropping the
  assumption that the first element of the orthogonal polynomial
  sequence be a constant. This approach gives rise to new families of
  complete orthogonal polynomial systems that arise as solutions of
  second-order eigenvalue equations with rational coefficients. The
  results are based on a classification of \emph{exceptional
    polynomial subspaces} of codimension one under projective
  transformations.
\end{abstract}

\maketitle

\maketitle

%
\section{INTRODUCTION AND STATEMENT OF RESULTS}
\label{sect:intro}
A classical question in the theory of linear ordinary differential
equations, which goes back to E. Heine \cite{He}, and which is at
the source of many important developments in the study of orthogonal
polynomials, is the following: given positive integers $m$ and $n$
and polynomials $p(x)$ and $q(x)$ with
\[
\deg p = m+2,\quad \deg q=m+1\,,
\]
find all the polynomials $r(x)$ of degree $m$ such that the ordinary
differential equation
\begin{equation}\label{Heineq}
p(x)y''+q(x)y'+r(x)y=0\,,
\end{equation}
has a polynomial solution of degree $n$. If there exists a
polynomial $r(x)$ solving Heine's problem, then it can be shown
\cite{Sz} that for that choice of $r(x)$ the polynomial solution $y$
of (\ref{Heineq}) is unique up to multiplication by a non-zero real
constant. Furthermore, it can also be shown that given polynomials
$p(x)$ and $q(x)$ as above, a sharp upper bound for the number of
polynomials $r(x)$ solving Heine's problem is given by
\begin{equation}\label{Heinebound}
\sigma_{nm}:=\binom{n+m}{n}\,.
\end{equation}

A well known interpretation of the bound (\ref{Heinebound}) is given
through an oscillation theorem of Stieltjes \cite{St}, which says
that if the roots of $p(x), q(x)$ are real, distinct, and
alternating with each other, then there are exactly $\sigma_{nm}$
polynomials $r(x)$ such that \eqref{Heineq} admits a polynomial
solution $y$ of degree $n$. Furthermore, the $n$ zeroes of these
solutions are distributed in all possible ways in the $m+1$
intervals defined by the $m+2$ zeroes of $p(x)$. This result also
admits a physical interpretation in the context of Van Vleck
potentials in electrostatics, where the roots of $y(x)$ are thought
of as charges located at the equilibrium configuration of the
corresponding Coulomb system.

The case $m=0$ is an important subcase of the Heine-Stieltjes
problem. The bound $\sigma_{n0}=1$ is exact; equation \eqref{Heineq}
is a variant of the hypergeometric equation that recovers the
classical orthogonal polynomials as solutions of \eqref{Heineq}
indexed by the degree $n$.  In this context, a related classical
question, posed and solved by Bochner \cite{Bo}, specializes the
Heine-Stieltjes equation \eqref{Heineq} to an eigenvalue problem.
\begin{thm}[Bochner]
  Let
  \begin{equation}
    \label{eq:Tydef}
    T(y) = p(x) y'' + q(x) y' + r(x) y
  \end{equation}
  be a second-order
  differential operator such that the eigenvalue problem
  \begin{equation}
    \label{eq:eigenproblem}
    T(P_n) = \lambda_n\, P_n\,,
  \end{equation}
  admits a polynomial solution $P_n(x)$, where $n=\deg P_n$, for
  every degree $n=0,1,2,\ldots$.  Then, necessarily the eigenvalue
  equation \eqref{eq:eigenproblem} is of Heine-Stieltjes type with
  $m=0$; i.e., the coefficients of $T$ are polynomial in $x$ with
  $\deg p = 2,\, \deg q=1,\, \deg r = 0$.
\end{thm}
\noindent If the above theorem is augmented by the assumption that the
sequence of polynomials $\{P_n(x)\}_{n\geq 0}$ is orthogonal relative
to a positive weight function, then the answer to Bochner's question
is given precisely by the classical orthogonal polynomial systems of
Hermite, Laguerre and Jacobi, as proved by Lesky, \cite{Lesky}.

\noindent \textbf{Remark 1.}
Although the literature in the past decades has referred to the result above as Bochner's theorem and still the name of Bochner is widely associated with the result, in recent years it has become clear that the question was already addressed  earlier by E. J. Routh, \cite{routh} (see page 509 of Ismail's book \cite{IsvAs}).

If we consider differential operators \eqref{eq:Tydef} with rational
coefficients, say
\begin{equation}
  p(x) = \tilde{p}(x)/s(x), \quad q(x) = \tilde{q}(x)/s(x),\quad
  \tilde{r}(x)/s(x),
\end{equation}
where $\tilde{p}, \tilde{q}, \tilde{r}, s$ are polynomials, then
the eigenvalue equation
\eqref{eq:eigenproblem} is, after clearing denominators, just a
special form of the Heine-Stieltjes equation \eqref{Heineq}, namely
\begin{equation}
  \label{eq:spectralheineq}
  \tilde{p}(x) y'' + \tilde{q}(x) y' + (\tilde{r}(x) - \lambda s(x)) y = 0.
\end{equation}
It is therefore natural to inquire whether it is possible to define
polynomial sequences as solutions to the Heine-Stieltjes equations
with $m>0$ ?  In the present paper, we show that this is possible by
weakening the assumptions of Bochner's theorem. Namely, we demand that
the polynomial sequence $\{ P_n(x)\}_{n=m}^\infty$ begins with a
polynomial of degree $m$, where $m>0$ is a fixed natural number,
rather than with a constant $P_0$.  If we also impose the condition
that the polynomial sequence be complete relative to some
positive-definite measure, then the answer yields new families of
orthogonal polynomial systems.

Let us consider the case $m=1$.  Let $b\neq c$ be constants, and let
\begin{equation}
  \label{eq:pxdef}
  p(x)=k_2 (x-b)^2+ k_1 (x-b) + k_0
\end{equation}
be a polynomial of degree $2$ or less, satisfying $k_0=p(b)\neq 0$.
Set
\begin{align}
  \label{eq:adef}
  a &= 1/(b-c),\\
  \label{eq:tqdef}
  \tilde{q}(x) &= a(x-c)(k_1 (x-b)+ 2k_0)\\
  \label{eq:trdef}
  \tilde{r}(x)&= -a(k_1 (x-b)+2 k_0)
\end{align}
and define the second-order operator
\begin{equation}
  \label{eq:TX1def}
  T(y):= p(x)y'' + \frac{\tilde{q}(x)}{x-b} y'
  +\frac{\tilde{r}(x)}{x-b} \, y,
\end{equation}
Observe that with $T$ as above, the eigenvalue equation
\eqref{eq:eigenproblem} is equivalent to an $m=1$ Heine-Stieltjes
equation:
\begin{equation}
  (x-b)p(x) y'' + \tilde{q}(x) y' +  (\tilde{r}(x)-\lambda (x-b)) y = 0
\end{equation}
We are now ready to state our extension of Bochner's result
\begin{thm}
  \label{thm:main1}
  Let $T$ be the operator defined in \eqref{eq:TX1def}. Then, the
  eigenvalue equation \eqref{eq:eigenproblem} defines a sequence of
  polynomials $\{ P_n(x)\}_{n=1}^\infty$ where $n=\deg P_n$ for every
  degree $n=1,2,3,\ldots$.  Conversely, suppose that $T$ is a
  second-order differential operator such that the eigenvalue equation
  \eqref{eq:eigenproblem} is satisfied by polynomials $P_n(x)$ for
  degrees $n=1,2,3,\ldots$, but not for $n=0$.  Then, up to an
  additive constant, $T$ has the form \eqref{eq:TX1def} subject to the
  conditions \eqref{eq:tqdef} \eqref{eq:trdef} and $p(b)\neq 0$.
\end{thm}

To put our result into perspective requires a point of view that is,
in some sense, the opposite of the one taken by Heine. Given a
collection of polynomials $y(x)$ we ask whether there exists a
$p(x)$ and a $q(x)$ such that this collection arises as the solution
set of a Heine-Stieltjes problem \eqref{Heineq}. Let
\begin{equation}
  \label{eq:cPdef}
  \cP_n(x) = \left< 1,x,\ldots, x^n \right>
\end{equation}
denote the vector space of univariate polynomials of degree less than
or equal to $n$.  Let $M=M_k\subset\cP_n$ denote a $k$-dimensional
polynomial subspace of fixed codimension $m = n+1-k$.  Let
$\cD_{2}(M)$ denote the vector space of second order linear
differential operators with rational coefficients preserving $M$.  The
assumption of rational coefficients is not a significant restriction.
Indeed, if $\dim M\geq 3$, then Proposition \ref{prop:ratcoeff} below,
shows there is no loss of generality in assuming that $\cD_2(M)$
consists of operators with rational coefficients.  We now arrive at
the following key definition.
\begin{definition}
  If $\cD_2(M)\not\subseteq \cD_{2}(\mathcal{P}_{n})$, we will call
  $M$ an \emph{exceptional} polynomial subspace. For brevity, we will
  denote by $X_m$ an exceptional subspace of codimension $m$.
\end{definition}
\noindent We will see below that the concept of an exceptional
subspace is the key ingredient that allows us to generalize
Bochner's result to a broader setting, and to thereby define new
sequences of polynomials as solutions of a second-order equation
\eqref{Heineq}. We now construct explicitly an $X_{1}$ subspace for
the operator \eqref{eq:TX1def} as a preparation for the proof of
Theorem \ref{thm:main1}.  With $a,b,c$ related by
\begin{equation}
  \label{eq:abcrel}
  a(b-c)=1
\end{equation}
we have
\begin{align}
  \label{eq:Taction1}
  T(x-c) &= 0 \\
   \label{eq:Taction2}
   T((x-b)^2) &= (2k_2 + a k_1)(x-b)^2+2 k_0a (x-c)\\
  \label{eq:Tactionn}
  T((x-b)^n) &= (n-1)(n k_2 + a k_1)(x-b)^n \\ \nonumber
  &\quad +  \big(n(n-2)k_1+2(n-1)ak_0\big)(x-b)^{n-1}\\ \nonumber
  &\quad +n(n-3)k_0(x-b)^{n-2},  \quad n\geq 2,
\end{align}
For $n=1,2,3,\ldots$, let
$\cE^{a,b}_n\subset\cP_n$ denote the following codimension 1
polynomial subspace:
\begin{align}
  \label{eq:Eabdef}
  \cE^{a,b}_n(x) &= \left< a(x-b)-1,(x-b)^2, \ldots,
    (x-b)^n\right>\\
  &= \left< x-c,(x-b)^2, \ldots,
    (x-b)^n\right>,\quad \text{if }  a\neq 0.
\end{align}
The above calculations show that $T$ leaves invariant the infinite
flag
\begin{equation}
  \cE^{a,b}_1 \subset \cE^{a,b}_2 \subset \cdots \subset \cE^{a,b}_n
  \subset \cdots,
\end{equation}
It is for this reason, that the eigenvalue equation
\eqref{eq:eigenproblem} defines a sequence of polynomials $P_1(x),
P_2(x), \ldots$.  By construction, each $P_n\in \cE^{a,b}_n$, while
equation \eqref{eq:Tactionn} gives the eigenvalues:
\begin{equation}
  \label{eq:eigenvalue}
  \lambda_n = (n-1)(nk_2+a k_1),\quad n\geq 1.
\end{equation}
Since $T$ has rational coefficients, it does not preserve $\cP_n$.
Hence, $T\in \cD_2(\cE^{a,b}_n)$ but $T\notin \cD_2(\cP_n)$, and
therefore $\cE^{a,b}_n$ is an $X_1$ subspace.  This observation is
responsible for the forward part of Theorem \ref{thm:main1}. A key
element in the proof of the converse implication (which we regard as an
extension of Bochner's theorem) is the following result, which states
that there is essentially one $X_{1}$ space up to projective
equivalence.
\begin{thm}
\label{thm:excep1} Let $M\subset\cP_n$ be an $X_1$ subspace.  If
$n\geq 5$, then $M$ is projectively equivalent to
  \[ \cE^{1,0}_n(x)  = \left<x+1,x^2,x^3,\ldots, x^n\right>.\]
\end{thm}
\noindent The answer appears to be much more restrictive than one
would have expected {\it a-priori}.  The notion of projective
equivalence of polynomial subspaces under the action of
$\SL(2,\mathbb{R})$, also an essential element of the proof, will be
defined at the beginning of Section \ref{Projeqsec}.  We complete
the proof of Theorem \ref{thm:main1} in Section
\ref{sect:proofconverse}.

One of the most important applications of Bochner's theorem relates to
the classical orthogonal polynomials.  In essence, the theorem states
that these classical families are the only systems of orthogonal
polynomials that can be defined as solutions of a second-order
eigenvalue problem.  However, new systems of orthogonal polynomials
defined by second-order equations arise if we drop the assumption that
the orthogonal polynomial system begins with a constant.

We are going to introduce two special families of orthogonal
polynomials that arise from flags of the form $\cE^{a,b}_n,
n=1,2,\ldots$ and that occupy a central position in the analysis of
the second order-differential operators that preserve codimension one
subspaces. The detailed analysis of these polynomial systems will be
postponed to a subsequent publication \cite{GKM5}.  Here we limit
ourselves to the key definitions and to the statement of our main
result concerning the $X_1$ orthogonal polynomials.

Let $\alpha\neq \beta$ be real numbers such that $\alpha,\beta>-1$
and such that $\operatorname{sgn}\alpha = \operatorname{sgn} \beta$.
Set
\begin{equation}
  \label{eq:jacobiabc}
  a= \frac{1}{2}(\beta-\alpha),\quad b=\frac{\beta+\alpha}{\beta-\alpha},\quad c = b+ 1/a.
\end{equation}
Note that, with the above assumptions, $|b|>1$. We define the
{\em{Jacobi-type $X_1$  polynomials}} $\hPab_n(x),\; n=1,2,\ldots$
to be the sequence of polynomials obtained by orthogonalizing the
sequence
\begin{equation}\label{eq:seq1}
 x-c, (x-b)^2, (x-b)^3,\ldots, (x-b)^n,\ldots
\end{equation}
 relative to the
positive-definite inner product
\begin{equation}
  \label{eq:jacobiproduct}
  \left< P,Q\right>_{\alpha,\beta} := \int^1_{-1} \frac{(1-x)^\alpha
  (1+x)^\beta}{(x-b)^2} P(x) Q(x) \, dx,
\end{equation}
and by imposing the normalization condition
\begin{equation}
  \label{eq:jacobinorm}
  \hPab_n(1) = \frac{\alpha+ n}{(\beta-\alpha)}
  \binom{\alpha+n-2}{n-1}.
\end{equation}
Having imposed \eqref{eq:jacobinorm} we obtain
\begin{equation}
  \label{eq:jacobiL2}
  \small
  \Vert \hPab_n \Vert^2_{\alpha,\beta} =
  \frac{(\alpha+n)(\beta+n)}{4(\alpha+n-1)(\beta+n-1)} C_{n-1},
\end{equation}
where
\begin{equation}
   C_n=\frac{2^{\alpha+\beta+1}}{(\alpha+\beta+2n+1)}
  \frac{\Gamma(\alpha+n+1) \Gamma(\beta+n+1)}{\Gamma(n+1)
    \Gamma(\alpha+\beta+n+1)}
\end{equation}
is the orthonormalization constant of $P^{(\alpha,\beta)}_n$, the
classical Jacobi polynomial of degree $n$.
\begin{prop}
  \label{prop:X1jacobi}
  Set $p(x) = x^2-1$ and let
  \begin{equation}
    \label{eq:JacobiTdef}
    T(y) = (x^2-1) y''+2a
    \left(\frac{1-b\,x}{b-x}\right) \big((x-c)y'-y\big),
  \end{equation}
  be the operator defined by equation \eqref{eq:TX1def}.  Then, the
  $X_1$ Jacobi polynomials $\hPab_n(x),\; n\geq 1$ form the solution
  solution set of the Sturm-Liouville problem given by \eqref{eq:eigenproblem}
  and boundary conditions
\begin{align}
      \label{eq:JacobiSLPa}
      &\lim_{x\to 1^-} (1-x)^{\alpha+1} (y(x)-(x-c)y'(x)) = 0,\\
      \label{eq:JacobiSLPb}
      &\lim_{x\to -1^+} (1+x)^{\beta+1} (y(x)-(x-c)y'(x)) = 0.
    \end{align}
    The corresponding eigenvalues are
  \begin{equation}
    \label{eq:jacobilambda}
    \lambda_n = (n-1)(n+\alpha+\beta)
  \end{equation}
\end{prop}

Likewise, for $\alpha>0$, we define the \emph{Laguerre-type $X_1$
  polynomials} to be the sequence of polynomials
$\hLa_n(x),\; n=1,2,\ldots$ obtained by orthogonalizing the sequence
\begin{equation}\label{eq:seq2}
 x+\alpha+1, (x+\alpha)^2, (x+\alpha)^3,\ldots, (x+\alpha)^n,\ldots
\end{equation}
relative to the
positive-definite inner product
\begin{equation}
  \label{eq:laguerreproduct}
  \left< P,Q\right>_{\alpha} := \int^\infty_0 \frac{e^{-x}
    x^\alpha}{(x+\alpha)^2}\, P(x) Q(x) \, dx,
\end{equation}
and normalized so that
\begin{equation}
  \label{eq:lagnorm}
  \hLa_n(x) = \frac{(-1)^nx^n}{(n-1)!} + \text{ lower order terms}.
\end{equation}

The orthonormalization constants are given by
\begin{equation}
  \label{eq:laguerreL2}
  \Vert \hLa_n \Vert^2_\alpha = \frac{\alpha+n}{\alpha+n-1} C_{n-1},
\end{equation}
where
\begin{equation}
  C_n = \frac{\Gamma(\alpha+n+1)}{n!}
\end{equation}
are the orthonormalization constants for $L^{(\alpha)}_n(x)$, the
classical Laguerre polynomial of degree $n$.

\begin{prop}
  \label{prop:X1laguerre}
  Set $p(x) = -x,\; a=-1,\; b=-\alpha$ and let
  \begin{equation}
    \label{eq:LaguerreTdef}
    T(y) = -x y'' + \frac{x-\alpha}{x+\alpha} \, ((x+\alpha+1)y'-y)
  \end{equation}
  be the operator defined by \eqref{eq:TX1def}.  Then, the $X_1$-Laguerre polynomials $\hLa_n$  form the solution set of the Sturm-Liouville problem
  defined by \eqref{eq:eigenproblem} and  boundary conditions
 \begin{align}
      &\lim_{x\to 0^+} x^{\alpha+1} e^{-x} (y(x)-(x-c) y'(x))=0,\\
      & \lim_{x\to \infty} x^{\alpha+1}e^{-x} (y(x)-(x-c)y'(x)) = 0.
    \end{align}
The corresponding eigenvalues are
  \begin{equation}
    \label{eq:laguerrelambda}
    \lambda_n = n-1.
  \end{equation}
\end{prop}

\noindent \textbf{Remark 2.}
Note that the weight factors \eqref{eq:jacobiproduct} and \eqref{eq:laguerreproduct} differ from the classical weights only by multiplication by a rational function. Ouvarov \cite{U} has shown how to relate via determinantal formulas the sequence of  polynomials obtained by Gram-Schmidt orthogonalization of the sequence $\{1,x,x^2,\dots\}$ with respect to two weights that differ by a rational function (see also Section 2.7 in \cite{IsvAs}). This does not mean however that Ouvarov's formulas apply to the $X_1$-Jacobi and $X_1$-Laguerre polynomials defined above, because although the weights differ by a rational function,\textit{ the two sequences to which Gram-Schmidt orthogonalization is applied are different}, i.e. they are $\{1,x,x^2,\dots\}$ for the classical polynomials but \eqref{eq:seq1} and \eqref{eq:seq2} for the $X_1$-polynomials.

Indeed, let $\{\tilde P_n\}_{n=0}^\infty$ be the sequence of polynomials obtained by Gram-Schmidt orthogonalization from the sequence $\{1,x,x^2,\dots\}$ with respect to the scalar product \eqref{eq:laguerreproduct}. Ouvarov's formulas relate the sequence $\{\tilde P_n\}_{n=0}^\infty$ with the classical Laguerre polynomials. However, the polynomials $\{\tilde P_n\}_{n=0}^\infty$ are semi-classical \cite{Ron87}: they do not satisfy a Sturm-Liouville problem, but only a second order differential equation whose coefficients depend explicitly on the degree of the polynomial eigenfunction. This is the case in general for rational modifications of classical weights and orthogonalization of the usual sequence, \cite{RM89}. By way of contrast, the $X_1$-Laguerre and $X_1$-Jacobi polynomials are eigenfunctions of a Sturm-Liouville problem as established by Propositions \ref{prop:X1jacobi} and \ref{prop:X1laguerre}.

Once this important precision has been made, we are now ready to state the following theorem, which is proved in \cite{GKM5}
\begin{thm}
  \label{thm:main}
  The Sturm-Liouville problems described in Propositions
  \ref{prop:X1jacobi} and \ref{prop:X1laguerre} are self-adjoint with
  a semi-bounded, pure-point spectrum. Their respective eigenfunctions
  are the $X_1$-Jacobi and $X_1$-Laguerre polynomials defined above.
  Conversely, if all the eigenfunctions of a self-adjoint, pure-point
  Sturm-Liouville problem form a polynomial sequence $\{ P_n
  \}_{n=1}^\infty$ with $\deg P_{n}=n$, then up to an affine
  transformation of the independent variable, the set of eigenfunctions is
  $X_1$-Jacobi, $X_1$-Laguerre or a classical orthogonal polynomial system.
\end{thm}

\noindent \textbf{Remark 3.}
In general, a classical orthogonal polynomial system $\{ P_n \}_{n=0}^\infty$ is no longer complete if the constant $P_0$ is removed from the sequence. However, in some very special cases the first few polynomials of the sequence (although solutions of the eigenvalue equation) do not belong to the corresponding $L^2$ space, while the remaining set is complete,\cite{elw04}. This happens for instance for Laguerre polynomials $L_n^\alpha(x)$ when $\alpha=-k$ is a negative integer: the truncated sequence $\{ L^{-k}_n\}_{n=k}^\infty$ forms a complete orthogonal basis of $L^2([0,\infty), x^{-k} {\rm e}^{-x}) \}$.

\noindent The new
polynomial systems described in Theorem \ref{thm:main} arise by
considering the $m=1$ case of the Heine-Stieltjes problem.  As was
noted above, this allows us to define a spectral problem based on the
flag of exceptional codimension 1 subspaces shown in
(\ref{eq:Eabdef}). This, in essence, is the ``forward'' implication
contained in Theorem \ref{thm:main}.  The reverse implication follows
from Theorem \ref{thm:main1}, but requires additional arguments that
characterize the $X_1$ Jacobi and Laguerre polynomials as the unique
$X_1$ families that form complete orthogonal polynomial
systems\footnote{Here, as part of the definition of an OPS, we assume
  that the inner product is derived from of a non-singular measure.}.
The proof of this result will be given in the following paper in this
series \cite{GKM5}.

Let us also point out that some $\X_1$ polynomial sequences can be
obtained from classical orthogonal polynomials by means of
state-adding Darboux
transformations~\cite{bagrov,dubov,GKM1}\footnote{The polynomials in
  question do not satisfy $\deg P_n = n$, but rather have $\deg P_1=0$
  and $\deg P_n =n$ for $n\geq 2$.  We will not consider them here.}
However, this does not explain the very restrictive answer that we
have obtained for what appears to be a rather significant weakening of
the hypotheses in Bochner's classification.  Let us also mention that
sequences of constrained, albeit incomplete, orthogonal polynomials
beginning with a first-degree polynomial have been studied in
\cite{Gi} as projections of classical orthogonal polynomials.

\section{THE EQUIVALENCE PROBLEM FOR CODIMENSION ONE SUBSPACES}
\label{Projeqsec} As a preliminary step to the proof of Theorem
\ref{thm:excep1} we describe the natural projective action of
$\SL(2,\mathbb{R})$ on $\cP_n$ and on the vector space of
second-order operators.  Our main objective here is to introduce a
covariant for the $\SL(2,\mathbb{R})$ action that will enable us to
classify the codimension one subspaces of $\cP_{n}$ up to projective
equivalence.

The irreducible $\SL(2,\mathbb{R})$ representation of interest here
is the following action, $P\mapsto \hP$, on $\cP_n$:
\begin{equation}
  \label{eq:Pnaction}
  \hP  = (\gamma \hx+\delta)^n\, (P\circ\zeta),\quad P \in   \cP_n,
\end{equation}
where
\begin{equation}
  \label{eq:fraclin}
   x=\zeta(\hx) = \frac{\alpha \hx + \beta}{\gamma \hx + \delta},\quad
  \alpha\delta - \beta\gamma =1
\end{equation}
is a fractional linear transformation.  The corresponding
transformation law for second-order operators is therefore given by:
\begin{equation}
  \label{eq:Taction}
  \hT(\hy) = (\gamma \hx + \delta)^n (T(y)\circ\zeta),
\end{equation}
where
\begin{equation}
  \label{eq:hydef}
   y(x) = (-\gamma x + \alpha)^n \hy\left(\frac{\delta x -
       \beta}{-\gamma x + \alpha}\right).
\end{equation}
Correspondingly, the components of the operator
undergo the following transformation:
\begin{align}
    \label{eq:sl2action-pqr}
    \hp  &=\ (\gamma \hx+\delta)^4 (p\circ \zeta),\\  \nonumber
    \hq &=  (\gamma \hx + \delta)^2 (q\circ \zeta)-2(n-
    1)\gamma(\gamma \hx + \delta)^3 (p\circ \zeta), \\ \nonumber
    \hr &=(r\circ\zeta)-n\gamma(\gamma \hx + \delta)  (q\circ\zeta)
     + n(n-1)\gamma^2 (\gamma \hx + \delta)^2 (p\circ\zeta).
\end{align}

For convenience, let us set the notation $V=\cP_n$ and
$G=\SL(2,\mathbb{R})$. Let $\cG_{n}(V)$ denote the Grassmann
manifold of codimension one subspaces of $V$, and let $\bP V =
\cG_1(V)$ denote $n$-dimensional projective space. We are interested
in the equivalence and classification problem for the $G$-action on
$\cG_{n}(V)$.  The action of $G$ is unimodular, and so there exists
a $G$-invariant $n+1$ multivector, which we denote by
$\omega\in\Lambda^{n+1}V$. Thus, we have a $G$-equivariant
isomorphism $\phi: \Lambda^{n}V\rightarrow V^*$, defined by
\begin{equation*}
  \phi(u_1\wedge \cdots \wedge u_{n})(u)\omega = u_1\wedge u_2\wedge
  \ldots \wedge u_{n}\wedge
  u,\quad u\in V.
\end{equation*}

Next, we define a non-degenerate bilinear form $\gamma:V\to V^*$ by
means of the following relations
\begin{equation*}
  n!\, \gamma\left(\frac{x^j}{j!},\frac{x^k}{k!}\right)= \begin{cases}
  \displaystyle (-1)^j & \text{, if }j+k=n,\\
  0 & \text{, otherwise.}
\end{cases}
\end{equation*}
Equivalently, we can write
\begin{equation}
  \label{eq:ginverse}
  \gamma^{-1}=\sum_{j=0}^n (-1)^{j}
  \binom{n}{j} x^j \otimes x^{n-j}.
\end{equation}
Note that $\gamma$ is symmetric if $n$ is even, and skew-symmetric if
$n$ is odd.
\begin{prop}
  The above-defined bilinear form  is $G$-invariant.
\end{prop}
\begin{proof}
Observe that
\begin{equation*}
  \Sym^2 V \cong\{ p(x,y)\in \Rset[x,y] : \deg_x(p)\leq
  n,\; \deg_y(p)\leq
  n\},
\end{equation*}
and that  the diagonal action of $G$ on  $\Sym^2 V$ is given by
\begin{equation*}
  \hp(\hx,\hy)= (\gamma \hx+\delta)^n (\gamma \hy+\delta)^n p\lp
  \frac{\alpha \hx+\beta }{\gamma \hx+\delta}, \frac{\alpha
    \hy+\beta}{\gamma \hy+\delta}\rp.
\end{equation*}
It is not hard to see that  $p(x,y) = (y-x)^n$ is an invariant.
Indeed,
\begin{equation*}
  \hp(\hx,\hy) = (\gamma \hx+\delta)^n (\gamma \hy+\delta)^n \left(
    \frac{\alpha
    \hy+\beta}{\gamma \hy+\delta}- \frac{\alpha \hx+\beta }{\gamma
    \hx+\delta}\right)^n = (\hy-\hx)^n.
\end{equation*}
Since,
\begin{equation}
  \label{eq:yxexpand}
  (y-x)^n = \sum_{j=0}^n (-1)^{n-j} \binom{n}{j} x^j y^{n-j}
\end{equation}
we see that $\gamma$ is invariant by comparing \eqref{eq:ginverse} and
\eqref{eq:yxexpand}.
\end{proof}

Since $\gamma$ is invariant, it follows that $\gamma^{-1}\circ\phi\colon
\Lambda^{n}V \longrightarrow V$ is a $G$-equivariant isomorphism.
This isomorphism descends to a $G$-equivariant isomorphism
$\Phi\colon \cG_n(V)\to \bP V$.
\begin{prop}
  Let $M\in\cG_n(V)$ be a codimension one subspace.  Then,
\begin{equation*}
\Phi(M) = \{ u\in V \colon \gamma(u,v)=0\text{ for all } v\in M\}.
\end{equation*}
\end{prop}
\noindent In other words, if $v_1,\ldots, v_n$ is a basis of $M$, we
can calculate $\Phi(M)$ by solving the $n$ linear equations
\begin{equation*}
  \gamma(v_j,u) = 0,\quad j=1,\ldots, n
\end{equation*}
for the unknown $u\in V$.

There is another natural way to exhibit the isomorphism between
$\cG_{n}(V)$ and $\bP V$.  Let $M\in \cG_n(V)$ be a codimension one
subspace with basis
$$p_i(x) = \sum_{j=0}^n p_{ij}\, x^j,\quad i=1,\ldots,n.$$ Let us now form
the polynomial \begin{equation}q_M(x) = \det
\begin{pmatrix}
  p_{10} & p_{11} & \ldots & p_{1j} & \ldots & p_{1n} \\
  p_{20} & p_{21} & \ldots & p_{2j} & \ldots & p_{2n} \\
   \vdots & \vdots & \ddots & \vdots & \ddots & \vdots \\
  p_{n0} & p_{n1} & \ldots & p_{nj} & \ldots & p_{nn} \\
  x^n & - n x^{n-1} & \ldots & (-1)^j \binom{n}{j} x^{n-j} & \ldots & (-1)^n
\end{pmatrix}
\end{equation}
The following proposition shows that, up to scalar multiple, this
polynomial characterizes $M$.
\begin{prop}
  With $q_M(x)$ as above, we have  $\Phi(M)= \left<q_M\right>$.
\end{prop}
\noindent Henceforth, we will refer to the subspace of $\cP_n$
spanned by $q_M$ as the fundamental covariant of the codimension one
subspace $M\subset \cP_n$.  Thanks to the $G$-equivariant
isomorphism between codimension one polynomial subspaces $M$ and
degree $n$ polynomials, we are able to classify the former by
considering the corresponding equivalence problem for degree $n$
polynomials. The latter classification problem can be fully solved
by means of root normalization, as one would expect.

Recall that a projective transformation \eqref{eq:fraclin} is fully
determined by the choice of images of $0, 1, \infty$.  Therefore, a
polynomial can be put into normal form by transforming the root of
highest multiplicity to infinity, the root of the next highest
multiplicity to zero, and the root of the third highest multiplicity
to 1.
\begin{prop}
  Every polynomial of degree $n$ is projectively equivalent to a
  polynomial of the form
  \begin{equation}
    \label{eq:rootnormalform}
    x^{n_0} (x-1)^{n_1} \prod_{j=2}^k  (x-r_j)^{n_j},
  \end{equation}
  where $r_j\neq 0,1,\; j\geq 2$ and where
  $$n=n_\infty+n_0+n_1+n_2+\ldots n_k,\qquad n_\infty\geq n_0\geq
  n_1\geq n_2\geq \cdots$$
  is an ordered partition of $n$.  The signature
partition $n_\infty, n_0, n_1,\ldots$ and the roots $r_j$ are
invariants that fully solve the equivalence problem.
\end{prop}
\noindent Note that in \eqref{eq:rootnormalform} there is no factor
corresponding to the multiplicity $n_\infty$; the missing factor
corresponds to the root at infinity.


It is instructive at this stage to show the expression of the
covariant $\left< q_M\right>$ of various codimension one subspaces
$M\subset \cP_n$.
\begin{enumerate}
\item Consider $M_1=\langle 1,x,\ldots,x^{n-1}\rangle\cong \cP_{n-1}(x)$.
  The fundamental covariant is $q_{M_1}(x) = 1$. In this case, $q_M$
  is equal to its own normal form; there is a single root of
  multiplicity $n$ at $\infty$.
\item Consider the exceptional monomial subspace:
  \[ M_2=\langle 1,x^2, x^3,\ldots, x^n\rangle= \cE^{0,0}_n(x).\]
  Section \ref{sect:laops} has more details on this example; see
  equation \eqref{eq:emod}.  The covariant in this case is $q_{M_2}(x)
  = x^{n-1}$.  The normal form of $q_{M_2}(x)$ is $x\in \cP_n(x)$;
  there is a root of multiplicity $n-1$ at $\infty$ and a simple root
  at $0$.
\item Consider the subspace \[M_3 = \langle
  1,x,x^2,\ldots,x^{n-2},x^n\rangle = \cE^0_n(x);\] see equation
  \eqref{eq:emod2}. In this case case $q_{M_3}(x)=x$.  Therefore,
  $M_2$ is projectively equivalent to $M_3$.  In section
  \ref{sect:laops}, below, we show that both $M_2$ and $M_3$ are $X_1$
  exceptional subspaces.
\item Consider a single gap monomial subspace,
  \[ M_4=\langle 1,x,\ldots ,x^{j-1},x^{j+1},\ldots, x^n\rangle.\] In this
  case, $q_{M_4}(x)=x^{n-j}$.  Here, the covariant has one root of
  multiplicity $j$ and another root of multiplicity $n-j$.
\end{enumerate}

In the next proposition, we classify the codimension 1 subspaces
$M\subset \cP_n$ directly, by exhibiting a normalized basis based on
the multiplicity of the root at infinity.
\begin{prop}
  \label{prop:Mbasis}
  Let $M\subset\cP_n$ be a codimension one polynomial subspace such that
  $q_M(x)$ has a root of multiplicity $\lambda$ at infinity and a root
  of multiplicity $\mu$ at zero; i.e., $\deg q_M=n-\lambda$ and $\mu$
  is the largest integer for which $x^{\mu}$ divides $q_M(x)$. The
  following monomials and binomials constitute a basis of $M$:
  \begin{equation}
    \label{eq:Mbasis}
    \{x^j\}_{j=0}^{\lambda-1},\quad  \{x^j+\beta_j x^\lambda\}_{j=\lambda+1}^{n-\mu}\quad \{ x^j
\}_{j=n-\mu+1}^n.
  \end{equation}
\end{prop}
\begin{proof}
  Observe that $q_M(x)$ has a root of multiplicity $\lambda$ at
  infinity and a root of multiplicity $\mu$ at zero if and only if, up
  to a scalar multiple,
  \begin{equation}
    \label{eq:qform}
    q_M(x) = (-1)^{\lambda}\binom{n}{\lambda} x^{n-\lambda} -
    \sum_{j=\lambda + 1}^{n-\mu} (-1)^j\binom{n}{j} \beta_j
    x^{n-j}.
  \end{equation}
  A straightforward calculation then shows that
  \[ \gamma(q_M,p)=0,\] where $p(x)$ ranges over the the monomials and
  binomials in \eqref{eq:Mbasis}.
\end{proof}

\section{OPERATORS PRESERVING POLYNOMIAL SUBSPACES}
\label{sect:laops} As was noted above, the standard $(n+1)$-dimensional irreducible representation of $\SL(2,\mathbb{R})$ can be
realized by means of fractional linear transformations, as per
\eqref{eq:Pnaction}.  The corresponding infinitesimal generators of
the $\mathfrak{sl}(2,\mathbb{R})$ Lie algebra are given by the
following first order operators
\begin{equation}
  \label{eq:sl2generators}
  T_- = D_{x},\quad T_0 = x D_{x} - \frac{n}{2},\quad T_+ = x^2
  D_{x} - nx.
\end{equation}
A direct calculation shows that the above operators leave invariant
$\cP_n(x)$, and are closed with respect to the Lie bracket:
\begin{equation}
  \label{eq:commutators}
  [T_0,T_{\pm}] = \pm T_{\pm},\quad [T_-,T_+] =  2 T_0.
\end{equation}
Since $\mathfrak{sl}(2,\mathbb{R})$ acts irreducibly on $\cP_n$,
Burnside's Theorem ensures that a second order operator $T$
preserves $\cP_n$ if and only if it is a quadratic element of the
enveloping algebra of the $\mathfrak{sl}(2,\mathbb{R})$ operators
shown in \eqref{eq:sl2generators}.
 Thus, the most general second order differential operator $T$
 that preserves $\cP_n$ can be written as
\begin{equation}
  \label{eq:liealgebraic}
  T = \sum_{i,j=\pm,\,0} c_{ij} T_i T_j  + \sum_{i=\pm,\,0} b_i T_i,
\end{equation}
where $c_{ij}=c_{ji}$, $b_i$ are real constants. For this reason, an operator that
preserves $\cP_n(z)$ is often referred to as a \emph{Lie-algebraic}
operator.

For the sake of concreteness we formulate results about invariant
polynomial subspaces
by assuming that all operators have  rational coefficients.  However, as the
following result will show, this assumption does not entail a loss of
generality.
\begin{prop}
  \label{prop:ratcoeff}
  Let $T$ be a second-order differential operator as per
  \eqref{eq:Tydef}.  Suppose that $P_i(x), Q_i(x), \; i=1,2,3$ are
  polynomials such that $P_1, P_2, P_3$ are linearly independent and
  such that $T(P_i) = Q_i$.  Then, necessarily the coefficients $p(x),
  q(x), r(x)$ of $T$ are rational functions.
\end{prop}
\begin{proof}
  By assumption,
  \[
  \begin{pmatrix}
    P_1''  & P_1' & P_1\\
    P_2''  & P_2' & P_2\\
    P_3''  & P_3' & P_3\\
  \end{pmatrix}
  \begin{pmatrix}
    p\\q\\r
  \end{pmatrix}
  =
  \begin{pmatrix}
    Q_1\\Q_2\\Q_3
  \end{pmatrix},
  \]
and the matrix on the left is non-singular.  Inverting
  this matrix, we obtain rational expressions for $p,q,r$.
\end{proof}

\begin{prop}
  \label{prop:labasis}
  A second-order operator $T$ preserves $\cP_n$ if and only if $T$ is
  a linear combination of the following nine operators:
  \begin{align}
    \label{eq:labasis1}
    &x^4 D_{xx} - 2(n-1)x^3 D_x+n(n-1)x^2,\\
    \label{eq:labasis2}
    &x^3 D_{xx} - 2(n-1)x^2 D_x + n(n-1) x,\\
    \label{eq:labasis3}
    &x^2 D_{xx} ,\; x D_{xx},\; D_{xx},\\
    \label{eq:labasis4}
    &x^2 D_x - n x,\\
    \label{eq:labasis5}
    &x D_x,\; D_x, 1
  \end{align}
\end{prop}
\noindent A proof can be given based on Burnside's theorem and
\eqref{eq:liealgebraic}.  For another proof, see Proposition 3.4 of
\cite{GKM4}.

Let us observe that Burnside's Theorem does not apply to general
polynomial subspaces $M\in \cP_n$, and therefore for a general
subspace $M$, there is no reason {\it{a priori}} for an operator
$T\in\cD_2(M)$ to also preserve $\cP_n$.  In addition to
\eqref{eq:Eabdef}, let us define the following codimension 1
subspaces:
\begin{align}
  \label{eq:Ebasis2}
  \cE_n^{a}(x) &=   \left< 1,x,x^2,\ldots ,x^{n-2},
  x^n  -  a x^{n-1}\right>.
\end{align}
Indeed, an analysis \cite{PT95, GKM3} of polynomial subspaces spanned
by monomials brought to light two special subspaces:
\begin{align}\label{eq:emod}
  \cE^{0,0}_n(x) &=
  \left<1,x^2,\dots,x^n\right>, \\ \label{eq:emod2}
  \cE^0_n(x) &= \left< 1,x,x^2,\dots,x^{n-2},x^n\right>.
\end{align}
These two subspaces  are
$\SL(2,\mathbb{R})$-equivalent, since
\begin{equation*}
  \cE^0_n(x) = x^n \cE^{0,0}_n(-1/x).
\end{equation*}
Extending the analysis beyond monomials we have the following
\begin{prop}
  \label{prop:X1subspaces}
  The subspaces $\cE^{a,b}_n, \cE^a_n$, as defined in
  \eqref{eq:Eabdef} and \eqref{eq:Ebasis2}, are all projectively
  equivalent.
\end{prop}
\noindent For the proof, see Proposition 4.3 of \cite{GKM4}.  Next,
we show that all of the above subspaces are $X_1$, that is
exceptional invariant subspaces of codimension one.
\begin{prop}
  \label{prop:Eop}
  A basis of $\cD_2(\cE_n^{a,b})$ is given by the following seven
  operators:
  \begin{align}
    \label{eq:jops1}
    J_1 &= (x-b)^4D_{xx} - 2(n-1)(x-b)^3D_x +n(n-1)(x-b)^2, \\
    \label{eq:jops2}
    J_2   &= (x-b)^3 D_{xx} -(n-1) (x-b)^2D_x\;,\\
    \label{eq:jops3}
    J_3 &= ( x-b)^2D_{xx} \;,\\
    \label{eq:jops4}
    J_4  & = ( x-b)D_{xx} + \lp a(x-b)-1\rp D_x \;,\\
    \label{eq:jops5}
    J_5 &=D_{xx}+2\lp a-\frac{1}{x-b}\rp \,D_x - \frac{2a}{
      x-b}\;,\\
    \label{eq:jops6}
    J_6 &= (x-b)\lp a(x-b)-n\rp D_x -an (x-b) ,\\
    \label{eq:jops7}
    J_7 &= 1.
  \end{align}
\end{prop}
\noindent The proof is given in Proposition 4.10 of \cite{GKM4}.

Observe that $J_5$ is an operator with rational coefficients. Hence,
$J_5$  preserves $\cE^{a,b}_n$, but does not preserve $\cP_n$.
Therefore, $\cE^{a,b}_n$ is an $X_1$ subspace. Because of projective
equivalence, so is $\cE^a_n$. Indeed, Theorem \ref{thm:excep1}
asserts that $\cE^{a,b}_n$ and $\cE^a_n$ are \emph{the only}
codimension one exceptional subspaces. We prove this theorem below.
In Section 3, we use Theorem \ref{thm:excep1} to establish Theorem
\ref{thm:main1}, our extension of Bochner's theorem.

\section{PROOF OF THEOREM \ref{thm:excep1}}\label{proof}
It will be useful to restate Theorem \ref{thm:excep1} in its
contrapositive form.
\begin{thm}
  \label{thm:allsl2}
  Let $M\subset \cP_n,\; n\geq 5$ be a codimension one subspace. If
  the roots of $q_M(x)$ have multiplicity less than or equal to $n-2$,
  then, $\cD_2(M)\subset \cD_2(\cP_n)$.
\end{thm}
\noindent In the preceding section, we showed that if $M$ is
projectively equivalent to $\cE^{0,0}_n$, then $q_M(x)$ has one root of
multiplicity $n-1$ and another root of multiplicity $1$.  On the other
hand, if $M$ is projectively equivalent to $\cP_{n-1}$, then $q_M$ has
a single root of multiplicity $n$.  Hence, if the roots of $q_M(x)$
have multiplicity less than or equal to $n-2$, then $M$ is not
isomorphic to  $\cE^{0,0}_n$ nor to $\cP_{n-1}$.  Theorem
\ref{thm:allsl2} asserts that, in this case, $\cD_2(M)\subset
\cD_2(\cP_n)$. The rest of the present section will be devoted to the
proof of this theorem.

We begin by writing a second-order differential operator with rational
coefficients using Laurent series:
\[ T=\sum_{k=-N}^\infty T_k \]
where
\begin{equation}
  \label{eq:Tkdef}
  T_k=x^k(a_k x^2 D_{xx} + b_k x D_x + c_k),\quad k\geq -N
\end{equation}
is a second-order operator of degree $k$, meaning that $T_k[x^j]$ is a
scalar multiple of $x^{j+k}$ for all integers $j$.  Henceforth, for a
series $L(x) =\sum_j L_j\, x^j$ we use the
notation
\[ \cf_j(L) = L_j.\]

Clearly, if $T$ is a differential operator such that $T(M)\subset M$,
then necessarily $T(M)\subset\cP_n$.  The converse, of course is not
true.  Nonetheless, it is useful to first classify all second order
operators that map $M$ into $\cP_n$, because in most instances this
larger class of operators turns out to preserve all of $\cP_n$.  To
complete the proof of the theorem, we consider the more restrictive
class of operators for which $T(M)\subset M$ for some limited cases.

The classification of operators $T$ which map $M$ to $\cP_n$ is the
subject of the subsequent lemmas.  Throughout the discussion, we
suppose that $T$ is a second-order differential operator and
$M\subset \cP_n$ is a codimension one subspace such that
$T(M)\subset \cP_n$. We also suppose that $q_M(x)$ has a root of
multiplicity $\lambda$ at
$\infty$, and a root of multiplicity
$\mu$ at $0$.
By Proposition \ref{prop:Mbasis}, this is equivalent to the
assumption that $x^j\in M$ for $j=0,\ldots,\lambda-1$, and
$j=n-\mu+1,\ldots, n$.
\begin{lem}
  If $T_k$ is an operator of fixed degree that annihilates three
  distinct monomials, that is if
  \[ T_k[x^j]=0 \] for three distinct $j$, then necessarily $T_k=0$.
\end{lem}
\begin{proof}
  Writing $T_k$ as in \eqref{eq:Tkdef} and applying it to $x^j$ gives
  \[ j(j-1) a_k + j b_k + c_k =0.\]
  Since the above equation holds for 3 distinct $j$, necessarily
  $a_k=b_k=c_k=0$.
\end{proof}
\begin{lem}
  \label{lem:1a}
  If $\lambda\geq 2$, then $T_k=0$ for all $\mid k\mid >n$.
\end{lem}
\begin{proof}
  By assumption, $1, x, x^n\in M$.  Hence, if $|k|>n$ the operator
  $T_k$ annihilates these monomials, and hence vanishes.
\end{proof}

\begin{lem}
  \label{lem:1b}
  If $q_M(x)$ has only simple roots, then $T_k=0$ for $\mid k\mid > n$.
\end{lem}
\begin{proof}
  By assumption, $T_k[1] = 0$ and $T_k[x^n]=0$ for all $|k|>n$.
  Hence,
  \begin{equation}
    \label{eq:1bak}
    T_k = a_k (x^{k+2} D_{xx} + (1-n) x^{k+1} D_x),\quad |k|>n.
  \end{equation}
  We are assuming $\mu=\lambda=1$, and hence, $x^j+\beta_j x\in M$ for
  $j=2,\ldots, n-1$.  This implies that
  \[ \cf_{k+1}(T[x^j+\beta_j x])=T_{k-j+1}[x^j] + \beta_j T_{k}[x] =0\]
  for all $k\geq n$ and all $k\leq -2$, and hence, by \eqref{eq:1bak},
  \begin{gather}
    \label{eq:1b2}
    j(j-n)a_{k-j+1} + (1-n)\beta_j a_{k} = 0,
  \end{gather}
  for all such $j$ and $k$.  In particular,  for $j=2$, we have
  \begin{equation*}
    a_{k-1} = \frac{n-1}{2(2-n)} \beta_2\, a_{k},
  \end{equation*}
  and more generally,
  \begin{equation}
    \label{eq:1b3}
    a_{k-j}=\left(\frac{n-1}{2(2-n)} \beta_2\right)^j a_{k}
  \end{equation}
  for all $j\leq k+1-n$ if $k\geq n$, and all $j\geq 0$ if $k\leq-2$.

  Let us argue by contradiction and suppose that $a_k\neq 0$ for some
  $k>n$ or for some $k<-n$.  By \eqref{eq:1b2} and \eqref{eq:1b3}, we have
  \begin{gather*}
    \beta_j = \frac{j(j-n)}{n-1}\left(\frac{n-1}{2(2-n)}
    \right)^{j-1} (\beta_2)^{j-1} ,\quad j=2,\ldots, n-1.
  \end{gather*}
  It follows that by setting
  \[  r= \frac{(n-1)}{2(2-n)} \beta_2 ,\]
  we have,   by \eqref{eq:qform},  that
  \begin{align*}
    q_M(x) &= -n x^{n-1} - \sum_{j=2}^{n-1} (-1)^j\binom{n}{j} \beta_j
    x^{n-j} \\
    &= -n x^{n-1} - \sum_{j=2}^{n-1} (-1)^j\binom{n}{j}
    \frac{j(j-n)}{n-1}\left(\frac{n-1}{2(2-n)} \right)^{j-1}
    (\beta_2)^{j-1}
    x^{n-j} \\
    &= -n x^{n-1} - \sum_{j=2}^{n-1} (-1)^j\binom{n}{j}
    \frac{j(j-n)}{n-1}r^{j-1}
    x^{n-j} \\
    &= -nx^{n-1} + n\sum_{j=2}^{n-1}
    (-1)^j\binom{n-2}{j-1} r^{j-1} x^{n-j} \\
    &= -nx (x-r)^{n-2}.\\
  \end{align*}
  This contradicts the assumption that all roots of $q_M(x)$ are
  simple.
\end{proof}

\begin{lem}
  \label{lem:2a}
  Suppose that $T_k=0$ for $k>n$.  If  $\lambda \leq n-3$, then,
  $T_k=0$ for $k\geq 3$, and
  \begin{align}
    \label{eq:T2}
    T_2 &= a_2 (x^4 D_{xx} +2(1-n) x^3 D_x + n(n-1) x^2) \\
    \label{eq:T1}
    T_1 &= a_1 x^3 D_{xx} + b_1 x^2 D_x - n((n-1)a_1  + b_1)x
  \end{align}
\end{lem}
\begin{proof}
  By assumption, $x^n, x^{n-1}+\beta_{n-1}x^\lambda, x^{n-2}+\beta_{n-2}
  x^\lambda\in M$; we do not exclude the possibility that
  $\beta_{n-1}=0$ or $\beta_{n-2}=0$. For $k\geq 3$,
  \begin{align*}
    &\cf_{k+n-1}(T[x^{n-1} + \beta_{n-1} x^\lambda]) = T_k[x^{n-1}] +
    \beta_{n-1} T_{k+n-1-\lambda}[x^\lambda] =0,\\
   &\cf_{k+n-2}(T[x^{n-2} + \beta_{n-2} x^\lambda]) =  T_k[x^{n-2}] +
   \beta_{n-2} T_{k+n-2-\lambda}[x^\lambda] =0.
  \end{align*}
  By assumption $n-1-\lambda,n-2-\lambda\geq 1$.  Hence, for $k=n$,
  by the above equations and by assumption,
  \[ T_n[x^{n-1}] = T_n[x^{n-2}] = 0.\]
  As well,
    \begin{equation*}
    T_k[x^n]=0,\quad k\geq 1.
  \end{equation*}
  Hence, $T_n$ annihilates three monomials, and therefore vanishes.  We
  repeat this argument inductively to conclude that $T_k=0$ for all
  $k\geq 3$. For $k=2$, we have
  \begin{equation*}
    T_2[x^{n-1}]=0,\quad T_2[x^n] =0,
  \end{equation*}
  and hence $T_2$ has the form shown in \eqref{eq:T2}.  Equation
  \eqref{eq:T1} follows from that fact that $T_1[x^n]=0$.
\end{proof}

\begin{lem}
  \label{lem:2b}
  Suppose that $T_k=0$ for $k>n$.  If $\lambda=n-2$, then, $T_k=0$ for
  $k\geq 4$, and
  \begin{align}
    \label{eq:T3a}
    T_3  &= a_3 (x^5 D_{xx} +2(1-n) x^4 D_x + n(n-1) x^3) \\    \label{eq:T2a}
    T_2 &= a_2 (x^4 D_{xx} +2(1-n)x^3
    D_x + n(n-1) x^2) + \\ \nonumber
    &\hskip 2em + 2\beta_{n-1} a_3 (x^3 D_x -n x^2)\\   \label{eq:T1a}
    T_1 &= a_1 x^3 D_{xx} + b_1 x^2 D_x - n((n-1)a_1  + b_1)x
  \end{align}
\end{lem}
\begin{proof}
  By assumption, $x^n,\, x^{n-1}+\beta_{n-1} x^{n-2},\, x^{n-3}\in M$;
  we do not exclude the possibility that $\beta_{n-1} = 0$.  Hence,
  for $k\geq4$,
  \begin{equation*}
    T_k[x^n]=0,\quad  T_k[x^{n-1}]+\beta_{n-1} T_{k+1}[x^{n-2}]=0,\quad
    T_k[x^{n-3}]=0 .
  \end{equation*}
  Since $T_{n+1}=0$, the above relations imply that $T_n$ annihilates
  three monomials, and hence vanishes. As before, we repeat this argument
  inductively to prove
  that $T_k=0$ for all $k\geq 4$.  For $k=3$, we have
  \begin{equation*}
    T_3[x^{n-1}]=0,\quad T_3[x^n]=0,
  \end{equation*}
  and hence  \eqref{eq:T3a} holds.
  As well,
  \[ T_2[x^n]=0,\quad T_2[x^{n-1}] + \beta_{n-1} T_3[x^{n-2}] = 0,\]
  which proves \eqref{eq:T2a}.  Finally, $T_1[x^n]=0$, which proves
  \eqref{eq:T1a}.
\end{proof}

\begin{lem}
  \label{lem:3b}
  Suppose that $T_k=0$ for $k<-n$.  If $\lambda\geq 3$, then $T_k=0$
  for $k\leq -3$, and
  \begin{align}
    \label{eq:3aT-2}
    T_{-2} &= a_{-2} D_{xx} \\
    \label{eq:3aT-1}
    T_{-1} &= a_{-1} x D_{xx} + b_{-1} D_x.
  \end{align}
\end{lem}
\begin{proof}
  By assumption, $1,x,x^2\in M$.  Hence, for all $k\leq -3$ the operator
  $T_k$ annihilates 3 monomials, and hence vanishes.  Also note that
  $T_{-2}$ annihilates $1,x$ and that $T_{-1}$ annihilates $1$.
  Equations \eqref{eq:3aT-2} \eqref{eq:3aT-1} follow.
\end{proof}

\begin{lem}
  \label{lem:3c}
  Suppose that $T_k=0$ for $k<-n$.  If $\lambda=2$ and $\mu \leq 2$,
  then the conclusions of Lemma \ref{lem:3b} hold.
\end{lem}
\begin{proof}
  By assumption, $1,x\in M$, and hence
  \begin{equation}
    \label{eq:3c1}
    T_k[1] = 0,\quad T_k[x] =0,\quad k\leq -2.
  \end{equation}
  As well, $x^{n-\mu}+\beta_{n-\mu} x^2\in M$, with
  $\beta_{n-\mu}\neq 0$, and hence, for $k\leq -3$,
  \begin{equation*}
    \cf_{k+2}(T[x^{n-\mu} + \beta_{n-\mu} x^2]) =
    T_{k+2-n+\mu}[x^{n-\mu}] + \beta_{n-\mu}T_{k}[x^2]=0.
  \end{equation*}
  If for some particular $k\leq -3$ we have that $T_{k+2-n+\mu}=0$,  then
  $T_{k}$  annihilates $1,x,x^2$. Hence,  by induction, $T_k=0$ for
  all $k\leq -3$.
\end{proof}

\begin{lem}
  \label{lem:3a}
  Suppose that $T_k=0$ for $k<-n$.  If $\mu=\lambda=1$, then then the
  conclusions of Lemma \ref{lem:3b} hold.
\end{lem}
\begin{proof}
  Since $\lambda=1$, we have $x^{n-1}+\beta_{n-1} x\in M$, where
  $\beta_{n-1}\neq 0$.  Hence, for $k\leq -3$, we have
  \begin{equation}
    \label{eq:3a1}
    \cf_{k+2}(T[x^{n-1}+\beta_{n-1} x])=T_{k+3-n}[x^{n-1}] + \beta_{n-1}
    T_{k+1}[x] =0,
  \end{equation}
  Since $\mu=1$, we have $x^2+\beta_2 x\in M$, and
  hence,
  \begin{equation}
    \label{eq:3a2}
    \cf_{k+2}(T[x^2]) = T_{k}[x^2] + \beta_2 T_{k+1}[x] =0.
  \end{equation}
  Arguing by induction, suppose that for a given $k\leq -3$, it has been
  shown that $T_j=0$ for all $j<k$ and that $T_k[x]=0$.  Since
  $\beta_{n-1}\neq 0$, \eqref{eq:3a1} implies that $T_{k+1}[x]=0$.
  Hence, by \eqref{eq:3a2}, $T_{k}[x^2]=0$, as well.  Since $1\in M$,
  we have
  \[ \cf_{k}(T[1]) = T_{k}[1]=0. \] Hence, $T_{k}=0$.  Our inductive
  hypothesis is certainly true for $k=-n$, and therefore it is true
  for all $k\leq -3$.  Furthermore, $T_{-2}[x]=0$. Since
  $T_{-2}[1]=0$, as well, \eqref{eq:3aT-2} follows.  Relation
  \eqref{eq:3aT-1} follows from the fact that $T_{-1}$ annihilates 1.
\end{proof}

\begin{proof}[Proof of Theorem \ref{thm:allsl2}]
  Let $M\subset\cP_n$ be a codimension 1 subspace with fundamental
  covariant $q_M(x)$. Let $T$ be a second-order operator such that
  $T(M)\subset M$.  Necessarily, $T(M)\subset\cP_n$, and so we can
  apply the above lemmas.  Let $\lambda$ be the maximum of the
  multiplicities of the roots of $q_M(X)$.  We perform an
  $\SL(2,\mathbb{R})$ transformation \eqref{eq:fraclin} so as to move
  the root of $q_M(x)$ with multiplicity $\lambda$ to $\infty$.  Since
  we have assumed that $q_M$ has at least two distinct roots, we may
  simultaneously move one of the other roots to zero.  Thus, without
  loss of generality, we suppose that $\infty$ and $0$ are roots of
  $q_M(x)$ with multiplicities $\lambda$ and $\mu\leq \lambda\leq n-2$,
  respectively, and that the multiplicity of all roots of $q_M(x)$ is
  $\leq \lambda$.

  Lemmas \ref{lem:1a} and \ref{lem:1b} establish that $T_k=0$ for
  $|k|>n$.  Next, we establish that $T_k=0$ for $k\geq 3$ and that
  $T_1, T_2\in \cD_2(\cP_n)$.   Here there are two cases to consider
  \begin{itemize}
  \item[(1)] If   $\lambda\leq n-3$, then Lemma
    \ref{lem:2a} establishes the above claims.
  \item[(2)]  Suppose that $\lambda = n-2$.
    Then, $1,x,\ldots, x^{n-3},
    x^{n-1}+\beta_{n-1}x^{n-2}, x^n$ is a basis for $M$; we do not
    exclude the possibility $\beta_{n-1}=0$. Since $T[x^{n-4}] \in M$,
    we have
    \begin{equation*}
      \beta_{n-1}\cf_{n-1}(T[x^{n-4}]) = \cf_{n-2}(T[x^{n-4}]),
    \end{equation*}
    which, by Lemma \ref{lem:2b}, is equivalent to
    \begin{equation*}
      12\beta_{n-1} a_3 = 12 a_2 -8 \beta_{n-1} a_3
    \end{equation*}
    Since $T[x^{n-5}]\in M$, we have
    \begin{equation*}
      20 a_3 = 0.
    \end{equation*}
    Therefore,  $T_k=0$ for $k\geq 4$, by Lemma
    \ref{lem:2b}. The above arguments establish that $a_3=0$.
    Therefore, by equations \eqref{eq:T3a} \eqref{eq:T2a}
    \eqref{eq:T1a}, $T_3=0$ and $T_2, T_1\in \cD_2(\cP_n)$.
  \end{itemize}

  Next, Lemmas \ref{lem:3b}, \ref{lem:3c}, \ref{lem:3a} establish that
  $T_k=0$ for $k\leq -3$, and that $T_{-2}, T_{-1}\in\cD_2(\cP_n)$.
  Finally $T_{0}\in \cD_2(\cP_n)$ by inspection.  Therefore,
  \begin{equation*}
    T=\sum_{k=-2}^2 T_k
  \end{equation*}
  is a sum of operators that preserve $\cP_n$ and therefore preserves
  $\cP_n$ itself.
\end{proof}

\section{PROOF OF THEOREM \ref{thm:main1} }
\label{sect:proofconverse} As it was noted in Section \ref{sect:intro},
the forward implication of Theorem \ref{thm:main1} is established by
equations \eqref{eq:Taction1} \eqref{eq:Tactionn}.  Here we prove the
converse.  Thus we suppose that $T$ is a second-order differential
operator with rational coefficients such that the eigenvalue equation
\eqref{eq:eigenproblem} has polynomial solutions $P_n(x)$ of degree
$n$ for integers $n\geq 1$, but not for $n=0$.  Set
\[ M_n = \langle P_1, P_2,\ldots, P_n \rangle,\quad n\geq 1.\] By
assumption, each $M_n$ is a codimension 1 subspace. By Theorem
\ref{thm:excep1}, for every $n\geq 5$, either $T$ preserves $\cP_n$,
or $M_n\cong \cE^{1,0}_n$.

Suppose that $T\in \cD_2(\cP_n)$ for some $n\geq 5$.  By Proposition
\ref{prop:labasis}, $T$ is a linear combination of operators
\eqref{eq:labasis1} - \eqref{eq:labasis4}.  However, since $T$ also
preserves $M_{n+1}$ and $M_{n+2}$, and since the operators
\eqref{eq:labasis1} \eqref{eq:labasis2} \eqref{eq:labasis4} have an
explicit dependence on $n$, our operator $T$ must be of the form
\[ T(y) = p(x) y'' + q(x) y' + r y,\]
where $\deg p = 2, \deg q = 1$ and where $r$ is a constant.
However, such an operator satisfies the eigenvalue equation
\eqref{eq:eigenproblem} for $n=0$, and hence can be excluded by
assumption.

Therefore, $M_n \cong \cE^{1,0}_n$ for all $n\geq 5$.  Proposition
\ref{prop:X1subspaces}, asserts that for $n\geq 5$, there exist
constants $a_n,b_n$ such that $M_n$ is either $\cE^{a_n,b_n}_n$ or
$\cE^{a_n}_n$, as per \eqref{eq:Eabdef} \eqref{eq:Ebasis2}.  We can
rule out the latter possibility, because by assumption, $M_n$ does not
contain any constants.  Hence, $M_n=\cE^{a_n,b_n}_n$, where, for the
same reason, $a_n\neq 0$.  Hence, there exist constants $b_n,c_n$ such
that
\[ M_n = \langle x-c_n, (x-b_n)^2,\ldots, (x-b_n)^n\rangle,\quad n\geq
5.\]
However, $x-c_5$ and $x-c_n$ are both a multiple of $P_1(x)$, and
hence $c_n=c_5$.   Also observe that  every polynomial $p\in M_n$
satisfies
\[  (c_n-b_n)p'(b_n) + p(b_n) = 0.\]
However, since $P_1, P_2, P_3$ also satisfy
\[ (c_5-b_5) y'(b_5) +  y(b_5) = 0,\]
we can apply the above constraint to $y(x) = (x-b_n)^2$ and $y(x)=
(x-b_n)^3$ to obtain
\begin{align*}
  & 2 (c_5-b_5)(b_5-b_n) + (b_5-b_n)^2 = 0,\\
  & 3 (c_5-b_5)(b_5-b_n)^2 + (b_5-b_n)^3 = 0.
\end{align*}
The above imply that $b_n = b_5$ also.  Henceforth, let us set
$b=b_5=b_n$, $c=c_5 =c_n$, $a = 1/(c-b)$.  We have established that
for every $n$,
\[ M_n = \cE^{a,b}_n(x) = \langle x-c, (x-b)^2,\ldots,
(x-b)^n\rangle.\] Hence, by Proposition \ref{prop:Eop}, $T$ is a
linear combination of the operators \eqref{eq:jops1} -
\eqref{eq:jops7}.  Again, operators $J_1, J_2, J_6$ have an explicit
dependence on $n$, and hence, up to a choice of additive constant, $T$
must be have the form
\begin{align*}
  T(y) &= (k_2 J_3 + k_1 J_4+ k_0 J_5-a k_1 J_7)(y)\\
  &= (k_2 (x-b)^2 + k_1 (x-b) + k_0 )y'' +\\
  &\quad +a(k_1+2k_0/(x-b))
  ((x-c)y'-y).
\end{align*}
By assumption, $T(1)$ is not a constant. Hence, by setting
\begin{align*}
  p(x) &= k_2 (x-b)^2 + k_1 (x-b) + k_0,
\end{align*}
we demonstrate that, up to an additive constant, $T$ has the form
\eqref{eq:TX1def} subject to the condition $p(b)\neq 0$.  This
establishes the reverse implication of Theorem \ref{thm:main1}.

\bigskip
\paragraph{\textbf{Acknowledgments}}
\thanks{
We thank Peter Crooks for reviewing the manuscript and providing many useful remarks.
 We are also grateful to Jorge Arves\'u, Norrie Everitt, Mourad Ismail, Francisco
  Marcell\'an, Lance Littlejohn and Andr\'e Ronveaux for their helpful comments on the manuscript.
This work was supported in part by the Ram\'on y Cajal program of the
Spanish Ministry of Science and Technology; the Direcci\'on General de Investigaci\'on [grants MTM2006-00478
and MTM2006-14603 to D.G.U.] and the National Science and Engineering Reserach Council of Canada [grants RGPIN 105490-2004 to N.K. and RGPIN-228057-2004 to R.M.]
}

\end{document}